\documentclass{article}

\usepackage{graphicx}  
\usepackage{wrapfig}
\usepackage{xcolor}
\usepackage{soul}

\PassOptionsToPackage{numbers, compress}{natbib}
\usepackage[final]{nips_2018}

\usepackage[utf8]{inputenc} 
\usepackage[T1]{fontenc}    
\usepackage{hyperref}       
\usepackage{url}            
\usepackage{booktabs}       
\usepackage{amsfonts}       
\usepackage{amsmath}
\usepackage{microtype}      
\usepackage{multirow}

\usepackage{subcaption}
\usepackage{cleveref}
\usepackage{authblk}

\newcommand{\ppn}{\textsc{PPN}}
\newcommand{\pptn}{\textsc{PPTN}}
\newcommand{\ppnlong}{{Phosphonate}}
\newcommand{\pptnlong}{{Phosphonothionate}}

\graphicspath{{images/}} 

\title{Chemical Structure Elucidation from Mass Spectrometry by Matching Substructures}

\author{Jing Lim\thanks{Indicates equal contributions. This work was conducted when Jing Lim served as a research intern in DSO National Laboratories.}~$\,^{2}$, Joshua Wong$^{\ast1}$, Minn Xuan Wong$^{\ast1}$, Lee Han Eric Tan$^{1}$, Hai  Leong Chieu$^{1}$, Davin Choo$^{3}$, Neng Kai Nigel  Neo$^{4}$\\
$^1$~DSO National Laboratories, Singapore,
$^2$~Department of Computer Science, Stanford University,
$^3$~ETH, Zurich, 
$^4$~Department of Chemistry, National University of Singapore  \\
{\tt\small jinglim2@stanford.edu, \{wjialain,wminnxua,tleehan,chaileon\}@dso.org.sg, chood@ethz.ch, nnnk@u.nus.edu}
}

\begin{document}

\maketitle
\begin{abstract}
Chemical structure elucidation is a serious bottleneck in analytical chemistry today. We address the problem of identifying an unknown chemical threat given its mass spectrum and its chemical formula, a task which might take well trained chemists several days to complete. Given a chemical formula, there could be over a million possible candidate structures. We take a data driven approach to rank these structures by using neural networks to predict the presence of substructures given the mass spectrum, and matching these substructures to the candidate structures. Empirically, we evaluate our approach on a data set of chemical agents built for unknown chemical threat identification. We show that our substructure classifiers can attain over 90\% micro F1-score, and we can find the correct structure among the top 20 candidates in 88\% and 71\% of test cases for two compound classes. 
\end{abstract}

\section{Introduction}

The identification of unknown chemical compounds remains one of the major challenges for analytical chemistry. Continuous improvements to instrumental capabilities have allowed a much bigger number of compounds to be separated and detected even at low concentrations. As a result, many new compounds in environmental or biological matrices can now be found in a short time, rendering chemical structure elucidation (CSE) a serious bottleneck in the analysis workflow. This has resulted in a greater emphasis in computer-aided CSE for fast and accurate identification of unknown compounds. State-of-the-art commercial software built to assist CSE requires user interaction with an analyst, or additional chemical post filtering processes such as boiling point, retention index and partitioning coefficient~\cite{schymanski2008use,schymanski2011automated}. 

\begin{figure}
\centering
\begin{subfigure}{0.75\textwidth}
\includegraphics[width=\textwidth]{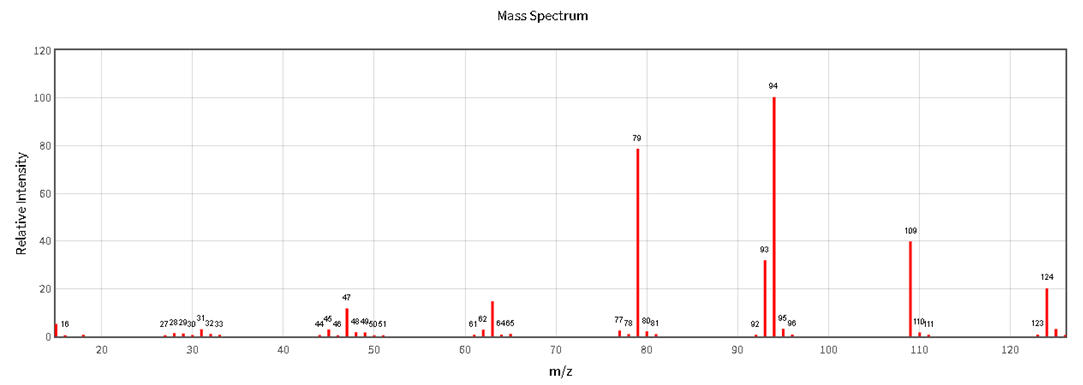}
\caption{Mass spectrum}
\label{fig:problema}
\end{subfigure}
\begin{subfigure}{0.22\textwidth}
\includegraphics[width=\textwidth]{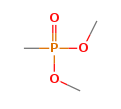}
\caption{Molecular structure}
\label{fig:problemb}
\end{subfigure}
\caption{
We show the mass spectrum and the corresponding molecular structure of dimethyl methylphosphonate ($C_3H_9O_3P$). This example is drawn from the NIST Chemistry Webbook~\cite{nist18}.}
\label{fig:problem}
\end{figure}

In this paper, we apply CSE to identify chemical agents controlled by the Chemical Weapons Convention~\cite{clinton93}. The use of structurally information-rich techniques such as nuclear magnetic resonance (NMR) require high sample purity and relatively high concentrations, and when these are not available, chemists could only obtain the mass spectrum and its chemical formula through the use of various gas chromatography detectors~\cite{schymanski2011automated}. A mass spectrum is a plot of the ion signal as a function of the mass-to-charge ratio, reflecting the relative abundance of detected ions. Usually the first strategy in CSE is to match the spectrum against a library of known spectra. For unknown compounds, a manual or software assisted interpretation of the mass spectra must be performed. This task is non trivial and might take a well trained chemist a few days, as a mass-to-charge ratio value (m/z) with only integer precision can represent an immense number of theoretically possible ion structures. We show an example of a mass spectrum and it's corresponding chemical structure in Figure~\ref{fig:problem}. Given the mass spectrum (e.g., Figure~\ref{fig:problema}) and the chemical formula (e.g.,  $C_3H_9O_3P$.), the objective of CSE is to pin down the chemical structure of the compound (e.g., Figure~\ref{fig:problemb}).  

Software such as MOLGEN~\cite{kerber2001molgen} could enumerate all possible structures given a chemical formula, but the number of possible structures could be in the millions. Previous work attempted to incorporate different substructure classifiers and filtering mechanisms, but the workflow always involve a chemist in the process to fine-tune substructure selection~\cite{kerber2001molgen,schymanski2008use,schymanski2011automated}. In this paper, we propose to address this problem in a completely automated workflow, presenting a list of ranked candidates as outputs for identification. We experimented on two compound classes in a data set provided by the Organisation for the Prohibition of Chemical Weapons (OPCW). Our approach ranked the provided ground truth structure in the top 20 positions in 88\% and 71\% of test cases for two different compound classes. If our approach can narrow down the list of candidates to 20 or less, we could then apply more computationally intensive computer simulation of fragmentation of these candidates for a final identification.

\section{Related Work}
\label{sect:rw}

One notable early work on computer-aided CSE from mass spectrometry (MS) is \cite{varmuza1996mass}. They applied principal component analysis, latent discriminant analysis and neural networks with one hidden layer to substructure classification. Empirically they analyzed 4 cases studies, manually selecting substructure classifiers, and used predicted substructures to assist analysts to reduce the number of candidates. These substructure classifiers were incorporated into the MOLGEN-MS~\cite{kerber2001molgen} software. In~\cite{kerber2001molgen}, they mentioned that the substructure classifiers ``typically require a user interaction for a critical check of classification results and adding additional structural information''.
More recently, Schymanski, Meinert, Meringer, and Brack~\cite{schymanski2008use,schymanski2011automated} combined MOLGEN-MS with additional substructure information from the NIST database~\cite{nist2008mass} as well as a number  of post-filtering criteria such as boiling point, retention index and partitioning coefficient. Out of 71 test cases, they found that the MOLGEN-MS and NIST classifiers can reduce the number of candidates to below 100 in 70\% of the cases. With the post-filtering criteria, they further reduce the number to 93\% below 100 and 70\% below 10. In the use of substructure classifiers in MOLGEN-MS,
Schymanski, Meringer, and Brack \cite{schymanski2011automated} have cautioned that the trial and error approach is required to select the substructure classifier and a chemist’s judgement is needed to manually review the classifier list prior to candidate generation to ensure that they do not restrict the candidate generation in undesired ways which could potentially prevent a fast identification.

Together with the burgeoning use of high-resolution, high-accuracy MS and tandem MS, alternative identification strategies have arisen that involve the use of online compound database (PubChem~\cite{bolton2008pubchem} and Chemspider~\cite{pence2010chemspider})  searching. These have spurred multiple analytical companies such as Thermo Fisher, Agilent Technologies, Waters and ACD Labs to develop commercial software, such as Mass Frontier (Thermo Fisher), Masshunter Profinder (Agilent), Progenesis Q1 (Waters) and ACD/MS Workbook Suite (ACD Labs), that make use of these online databases to yield substructure information based on the mass spectrum of the unknown compound. With this information, the chemist can then postulate probable structures on which the software would perform an in-silico fragmentation and from the fragments generated, a match value to the experimental mass spectrum is calculated. However, not only is there a limited amount of spectral databases available for high-resolution and tandem mass spectrometry, the requirement of a well trained chemist to elucidate the unknown structure does not make the structural elucidation process any less of a challenge. 

In this work, we propose to automate the entire process from substructure selection to candidate ranking. Previous work~\cite{varmuza1996mass} used the ToSim~\cite{tosim94} software to search for substructures in compounds to generate the ground truth for training substructure classifiers. In this work, we compile training data for substructure classification by checking if substructures are subgraphs of the chemical compound structures. 
Now, substructures responsible for certain m/z in a mass spectra may not be subgraphs due to rearrangement reactions. On the other hand, the ground truth for the actual spectral fragments responsible for m/z spikes in a mass spectrum are unavailable unless we employ software such as Mass Frontier, an expensive operation if we have to run it through all candidates during testing. Hence, we trained our substructure classifiers based on ground truth established through subgraph isomorphism (SGI). As a result, the substructure classifiers learn to predict the presence of substructures as subgraphs, rather than as actual spectral fragments. Empirically, we show that the substructure classifiers can achieve high micro F1-scores, and in our analysis, we illustrate with examples of substructures which may not coincide with spectral fragments, but are still useful in the CSE process.

\section{Substructure Matching for CSE}

\begin{figure}
\centering
\includegraphics[width=\textwidth]{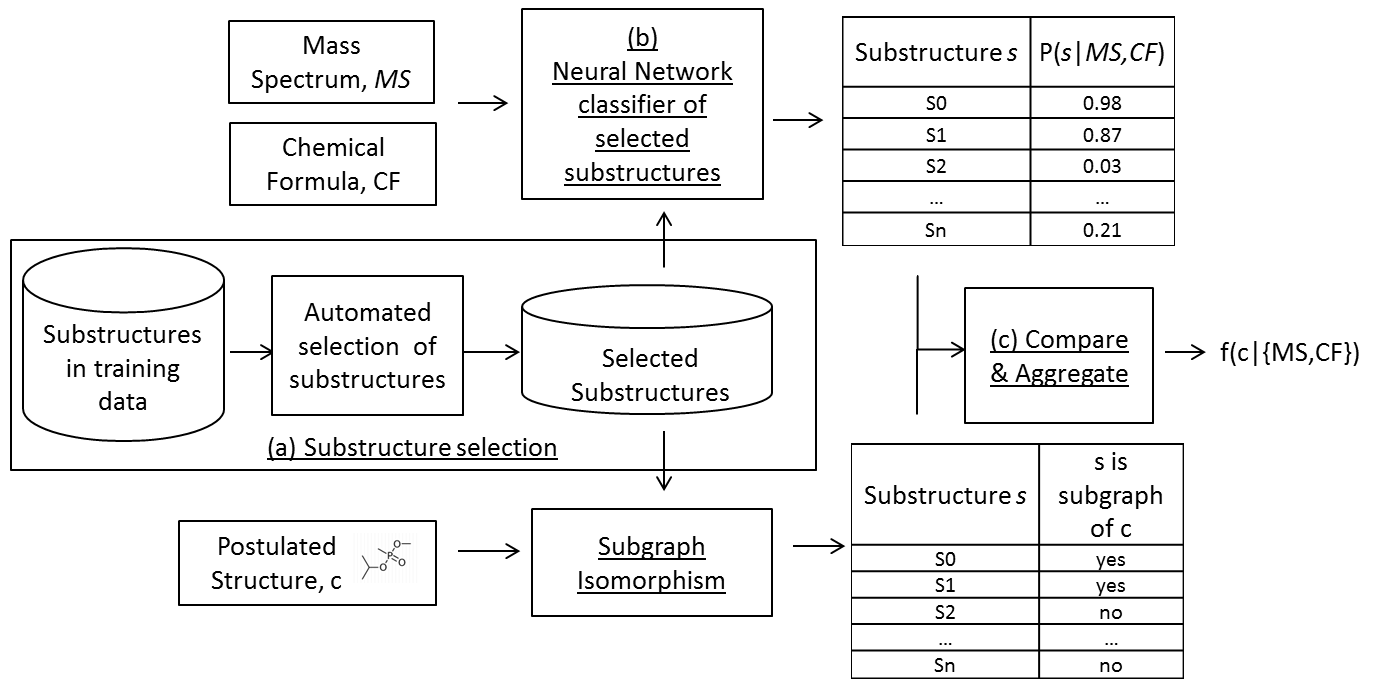}
\caption{The workflow for computing a score $f(c|\{MS,CF\})$ of a candidate structure, $c$, given a mass spectrum and its  chemical formula, $\{MS,CF\}$.}
\label{fig:workflow}
\end{figure}

We propose to elucidate the chemical structure of a compound given its mass spectrum and chemical formula. To rank candidate structures generated by MOLGEN, we build a scoring function that scores candidate structures against a given $(MS,CF)$ pair. We evaluate our results by the rank of the correct structure among the candidates. We illustrate the workflow for building the scoring function in Figure~\ref{fig:workflow}. Given a training data set of $(MS, CF)$ pairs mapped to their underlying molecular structure $S^*$, we first gather a set of substructures from the training data, and run an automated substructure selection algorithm to select a small number of useful substructures (box (a) in Figure~\ref{fig:workflow}). We then use machine learning to learn classifiers to predict the presence of the selected substructures given each $(MS, CF)$ pair (box b). During testing, we use these classifiers to compute the likelihood of each substructure given a $(MS,CF)$ pair. Given postulated candidate structures, we score each candidate by comparing predicted substructures against the candidate structure (box (c)).  We describe the components in more detail in the remaining of this section.

\subsection{Substructure selection}

From the training data of structures, we propose to first enumerate possible interesting substructures as candidate substructures for further selection. We tried the two approaches for enumerating possible substructures in a given structure. The first approach is a simple rule-based cleaving of bonds, and the second rely on a commercial software called Mass Frontier.

\paragraph{Three-stage fragmentation (3SF)}

The first approach is an iterative cleaving of bonds to obtain possible substructures: given structures in the training data, we gather the substructures resulting from cleaving any one edge, and perform this same cleaving action to each substructure recursively two more times.
The resulting substructures are counted, and we only keep those which occur within one standard deviation from the mean, discarding those that are either too common or too rare.

\paragraph{Mass Frontier (MF)}
For the second approach, we apply Mass Frontier from Thermo Fisher Scientific to generate stable substructures given a structure. We initialised MF through its fragmentation settings to generate fragments from molecules under electron impact ionisation. We also limit the number of reaction steps for fragment generation, ensuring that most of the fragments present in the spectrum are captured, but not unnecessarily extend the amount of computing time needed.
Similar to the previous substructure selection method, we only keep fragments whose occurrences are within one standard deviation from the mean.

\paragraph{Decision Tree Selection}
\label{sect:dts}

Denote the set of substructures (gathered by either MF or 3SF) as $CS$. In this section, we propose to select a small subset of $CS$ to be used for matching, for reasons of computational efficiency.
Since the purpose of the substructures is to help find the correct structure among all candidates generated by MOLGEN, we aim to select a subset of $CS$ that can help discriminate between candidates generated by MOLGEN. We take 1 to 3 representatives from the training data set and run them through MOLGEN to generate all possible candidate structures. We denote the set of generated structures as the set $C$. 
For each candidate $c \in C$, we run SGI to determine if each substructure $s \in CS$ is a subgraph of $c$, forming the vector $v(c) = \{\delta(s\in c)\}_{s\in CS}$, where $\delta(s\in c)$ is 1 if $s$ is a subgraph of $c$, and 0 otherwise.
We define the equivalence relation $c_1 \sim c_2$ if and only if $v(c_1) = v(c_2)$. Two equivalent candidates contain exactly the same substructures in $CS$ as subgraphs. 

The equivalence classes define the resolution power of the set $CS$: two equivalent candidates would not be distinguishable by matching substructures in $CS$. On the other hand, not all substructures in $CS$ are necessary to divide $C$ into these equivalence classes. We aim to select a minimal subset of $CS$ that is sufficient in dividing the candidates into these equivalence classes. 
We achieve this by running the decision tree classifier~\cite{breiman2017classification} in scikit-learn~\cite{sklearn11} by using the structures as instances, substructures as features, and equivalence classes as target labels. We collect all substructures with non-zero Gini-importance into $CS^*$, as our minimal subset of distinguishing substructures for matching.

In our experiments, for computation efficiency, we only used candidates generated from 1 to 3 representatives for  each compound class from the training data. Note that $CS^*$ has the same resolution power as $CS$ on the candidates used for selecting substructures, i.e. it divides $CS$ into the same equivalence classes. In our experiments, the MF approach of generating substructures resulted in around 100,000 unique substructures in $CS$, while $CS^*$ only contains around 1,000 substructures.

\subsection{Neural Network Classifier}
\label{sect:nnlearn}

\begin{figure}
\includegraphics[width=14cm]{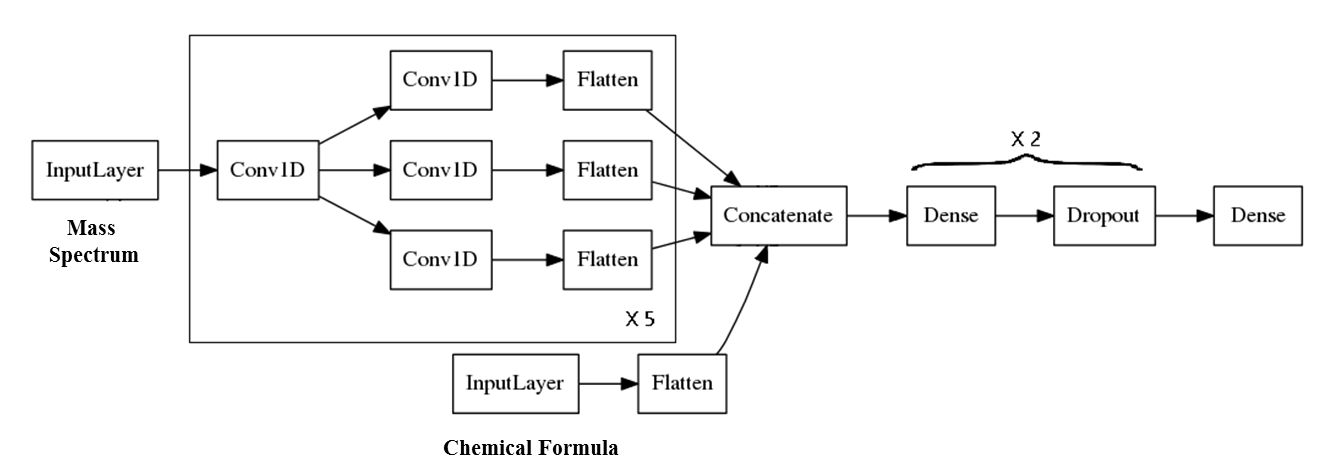}
\caption{Neural network architecture for predicting substructures.}
\label{fig:nnarch}
\end{figure}

In this section, we describe the neural network classifier used to predict the presence of substructures given $(MS,CF)$ pairs. The objective of the substructure classifier is to learn to predict the presence of a substructure in a compound, given clues from the mass spectrum. However, establishing the actual spectral fragments responsible for the spikes in the mass spectrum is too expensive during testing for matching candidates. Hence, we decided to train substructure classifiers to predict if a substructure is a subgraph of the structure that generates the mass spectrum.

Initial experiments with classifiers such as support vector machines~\cite{svm11} performed mediocrely for the final ranking evaluation. Hence, we built a one dimensional convolutional neural network (CNN) on the mass spectrum for feature learning, illustrated in Figure~\ref{fig:nnarch}. For an input $(MS, CF)$ pair, we L2-normalise $MS$ and represent $CF$ as counts of chemical elements existing in the compound.  We extract features from $MS$ using a two-layered convolutional neural network. The first convolution layer consists of filters of sizes 3, 4, 5, 6, and 7, with 100 filters for each sizes. The second convolution layer consists of filter sizes 2, 3, and 4, also with 100 filters for each size, resulting in a total of 150,000 filters. 
$CF$ is concatenated to the flattened outputs of the second CNN layer before classifying the input using two fully connected (FC) layers.
The number of dense units in each layer follows that of the number of output classes.
Between each FC layer, there is a dropout layer with dropout rate of 0.25.
The Rectified Linear Unit (ReLU) is used as activation functions except for the output layer which uses the sigmoid function.
The output are the probabilities $\{p(s|\{MF,CF\})\}_{s\in CS^*}$ of the substructures being present in the chemical compound given $\{MS,CF\}$. We train the neural networks for all selected substructures in a multi-task manner, where all the neural networks share all layers except the final layer.
We train the network for 100 epochs using Adam optimizer and minimise the binary cross entropy.

\subsection{Subgraph Isomorphism}

Subgraph isomorphism (SGI) is NP-complete [6]. As we use a large number of SGI computations, it is the main bottleneck in our workflow. To improve efficiency, we leverage the fact that, for a fixed compound, we compute SGI for a large number of similar graph and subgraph pairs.

\begin{wrapfigure}{R}{0.25\textwidth}
\centering
\includegraphics[width=0.2\textwidth]{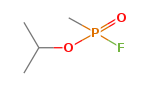}
\caption{Sarin's graph. Image from NIST Chemistry WebBook~\cite{nist18}.}
\label{fig:sarin}
\end{wrapfigure}

For any two graphs $G_i$ and $G_j$ such that $G_i$ is a subgraph of $G_j$ (notationally $G_i \subseteq G_j$), certain necessary conditions must hold. By contrapositive, if any of these conditions fail, then we can quickly deduce that $G_i$ is \emph{not} a subgraph of $G_j$. One such condition is the ``set of atoms''. For example, for $H_2 0$ to be a subgraph of $G_j$, $G_j$ \emph{must} contain at least 2 hydrogen atoms and 1 oxygen atom. For each graph and subgraph, we compute the following signatures: \textsc{atoms} refers to all atoms that appear in the graph, and \textsc{neighbours} is a list of the list of neighbours for each type of atom. As each substructure and candidate structure are involved in many SGI pair evaluation, the pre-processing time for graph signature computation is negligible compared to the actual time spent in SGI evaluations.

We illustrate the above-mentioned signatures with the Sarin compound $C_4 H_{10} F O_2 P$ (See figure \ref{fig:sarin}). We represent nodes by their atom string and edges by the number of bonds.

\begin{itemize}
    \item \textsc{atoms}(Sarin) = $\{C,C,C,C,F,O,O,P\}$
    \item \textsc{neighbours}(Sarin) = $\{Nbr[C], Nbr[F], Nbr[O] Nbr[P]\}$, where
    \begin{itemize}
        \item $Nbr[C] = \{\{C\},\{C\},\{P\},\{C,C,O\}\}$
        \item $Nbr[F] = \{\{P\}\}$
        \item $Nbr[O] = \{\{C,P\},\{P,P\}\}$
        \item $Nbr[P] = \{\{C,O,O,O,F\}\}$
    \end{itemize}
\end{itemize}

By comparing graph signatures, we only run the SGI implementation in igraph~\cite{igraph06} in cases where the signatures are insufficient in ruling out SGI.

\subsection{Final scoring and ranking}
\label{sect:score}

The final score of a candidate structure $C$ given a mass spectrum $MS$, and chemical formula, $CF$, is 
\begin{eqnarray}
f(c|\{MS,CF\}) = \prod_{s \in CS^*} \left(p(s|\{MS,CF\}) \cdot \delta(s \in c) + (1-p(s|\{MS,CF\})) \cdot \delta(s \notin c) \right),
\label{eqn:score}
\end{eqnarray}
where $CS^*$ is the set of substructures selected by the decision tree algorithm in Section~\ref{sect:dts}, and $p(s|\{MS,CF\})$ is output probability of the substructure neural network classifier. This is computed for all candidate structures generated by MOLGEN.  We then rank the candidates in descending order of their scores $f(c)$. Note that the $p(s|\{MS,CF\})$ in Equation~\ref{eqn:score} only depend on the $MS$ and not the candidate $c$, and so candidate structures which are in the same equivalence classes (i.e., contain exactly the same substructures in $CS^*$) would have identical scores, and hence will be ranked equally. Also note that $\delta(s \in c)$ is defined to be 1 if $s$ is a subgraph of $c$, and 0 otherwise. We decided to use SGI instead of checking if $s$ is a spectral fragment for $c$, as this check is expensive, and hence we have also trained the classifiers on ground truth established via SGI.

\section{Experimental Results}

We conducted experiments on the Validation Group Working Database (VGWD) 2017 provided by the Organisation for the Prohibition of Chemical Weapons (OPCW)~\cite{vgwd2017}. The database provided the chemical table files of all chemicals in the database as well as their corresponding spectra files. Using the index file given by OPCW, the organophosphorus compounds in the database were split into 29 compound classes according to their functional group. Due to the time needed to prepare and select the substructures for the training data, we decided to apply the workflow on each compound class separately, and we selected two compound classes for experimentation: \ppnlong{} (\ppn{}), the biggest compound class with 1715 instances, and \pptnlong{} (\pptn{}), the fourth in size with 293 instances, for our experiments.
 To show that the approach is general and not constrained by this limitation, we built a classifier that can classify organophosphorus compounds into the correct compound class that could serve as a pre-processing step. We conducted the compound class classification experiments on 27 of the 29 classes which contain more than 10 instances.  We show that we can obtain extremely high accuracies of over 99\% for this problem (see Section~\ref{sec:ccclass})

\begin{figure}
\centering
\includegraphics[width=\textwidth]{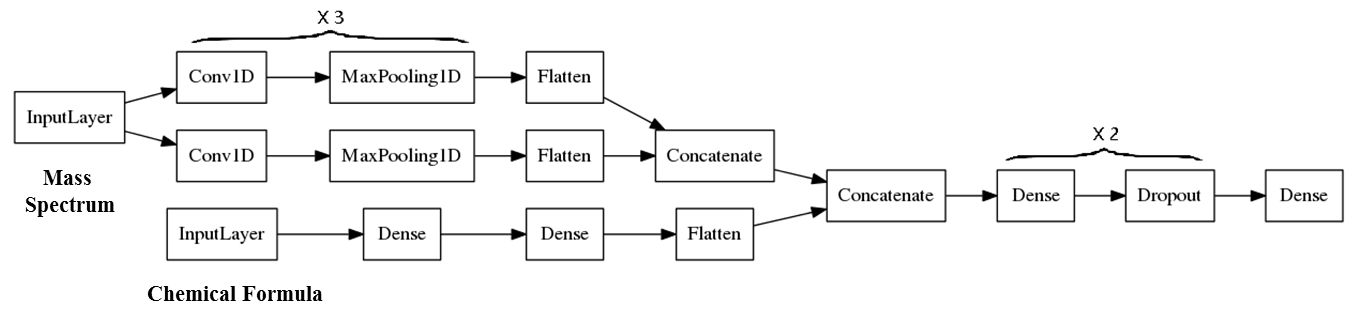}
\caption{Neural network architecture of compound class classifier.}
\label{fig:subclassifier}
\end{figure}

\subsection{Compound Class Classification}
\label{sec:ccclass}

We did a five-fold cross validation on the data set to classify compound classes into these 27 classes. The data set consists of a total of 4223 instances, with 12 in the smallest class and 1715 in the biggest class. This task seems a lot easier than the substructure classification task: simply training a support vector classifier~\cite{svm11} with default parameters gives an accuracy of 98.5\% with a linear kernel, 84.4\% with a polynomial kernel, and 84.8\% with a rbf kernel. 
We also tried with a neural network classifier similar to the one used for substructure classification, as shown in Figure~\ref{fig:subclassifier}. 
Features are extracted from the L2-normalized $MS$ using three one-dimensional convolution layers with filter sizes 3 and 4. The number of filters for each progressive layer are 64, 128 and 256. Between each convolution layer, max pooling is implemented with pool size and stride 2, to provide spatial invariance to overcome minor shifts in mass spectra peaks due to isotopes. Features are extracted from the $CF$ using two fully connected layers with 32 and 64 units respectively. Extracted features are flattened and concatenated and used as input to two fully connected layers with 128 units in each layer. Between the fully connected layers are dropout layers with dropout rate of 0.25. ReLU is used as activation function except the output layer where the softmax function is used instead.  The network was trained for 100 epochs and batch size of 32 using Adam optimizer and minimize the categorical cross entropy. We adopted the idea of smooth labelling commonly used in Generative Adversarial Networks~\cite{salimans2016improved} during training to reduce the extremity of confidence probability. We also trained with class weights (derived from the training data) to overcome class imbalance. We achieved 99.5\% accuracy with this neural network architecture.

\subsection{Data Preparation}

The OPCW VGWD 2017 database~\cite{vgwd2017} provides the chemical table files of all chemicals in the database as well as their corresponding spectra files, the latter being generated under electron impact condition. In order to obtain $CS$, the complete set of substructures, by the Mass Frontier approach, we run the Mass Frontier fragmentation on the whole training set, taking about 3 minutes per compound, and generating around 100,000 unique substructures. We experimented on \pptn{} (with 293 instances) and \ppn{} (with 1715 instances) compound classes, with random 5:1 train-test split ratio resulting in1430/285 and 244/49 compounds for the two classes respectively. In the substructure selection process with decision tree described in Section~\ref{sect:dts}, we used MOLGEN to generate candidates for representatives for each compound class. These representatives are chosen based on the different functional groups that each compound class would possess.  We list the representatives selected for our experiments in Table~\ref{tab:representatives}. In our experiments, we compare our results using different number  of representatives for substructure selection.

\begin{table}
\centering
\begin{tabular}{cccl} \toprule
Class & Id & \# Cand & Name \\  \midrule
\pptn{} & 1 	& 848       	& O-sec-butyl o-propyl ethylphosphonothionate \\
		& 2  	& 17758		& O-propyl o-trimethylsilyl isopropylphosphonothionate
 \\ \midrule
\ppn{} &  1	& 2092       	& 3-methylbutyl propyl ethylphosphonate\\
 &  2	& 48872         	& Pinacolyl trimethylsilyl methylphosphonate \\
 &  3	& 45522       	& Propyl 2-diethylaminoethyl methylphosphonate  \\
\bottomrule \\
\end{tabular}
\caption{Representatives used in the substructure  selection with decision trees for each compound class. The \#Cand column shows the number of structures generated by MOLGEN given the chemical formula.}
\label{tab:representatives}
\end{table}

We use MOLGEN to generate candidate structures for ranking during evaluation. However, MOLGEN can take very long to generate all candidates, and it will only produce any output at all at the end of its execution. For efficiency, we configured MOLGEN to stop at 1 million candidates.  MOLGEN can also be configured to generate only candidates that agree with specified annotations, which reduces the number of candidates generated. We conducted two sets of experiments as follows:
\begin{enumerate}
\item[ADDGT] We run MOLGEN without annotation. In cases where there are more than 1 million candidates and the ground truth structure is not among the 1 million candidates, we add the ground truth into the list of candidates. In this  experiment, we compare the performance of the two substructure generation approaches, 3SF and MF.  We also  assume that compound class classification is performed perfectly in this setting.
\item[STRICT] We constrain MOLGEN with annotations to the phosphorus (valency = 5, hydrogen bonds = 0, double bond = 1), sulfur  (double bonds = 1) and oxygen (double bonds = 1) atoms, depending on the compound class (\ppn{} or \pptn{}), so that only valid candidates of the class are generated. For the  family \pptn{}, MOLGEN failed to generate the ground truth in two instances (because we stopped the generation at 1 million), but for \ppn{}, MOLGEN failed in 41 instances. In these cases, we count these test instance as failures. Cases wrongly classified in compound  class classification are also counted as failures.
\end{enumerate}

\subsection{Results}

In this section, we report results using the neural networks described in Section~\ref{sect:nnlearn}. To compare 3SF and MF substructure generation methods, we evaluate them in the ADDGT settings and show the results in Table~\ref{tab:easy}. We see that the MF approach does slightly better than 3SF  for ranking the ground truth in the top 5 positions. The MF approach would produce more chemically viable substructures as it is based on chemical principles. However, the results we achieved with 3SF is only slightly worse, especially for the larger class \ppn{}.

\begin{table}
\centering
\small
\begin{tabular}{llcrrrrr}
\toprule
Class (\#cases)  & SG & \#REP & \#SSTR & \#CAND & Time & Top 5 & Top 20 \\
\midrule
\ppn{} (285) & 3SF & 1 & 1,131 & 641,032 & 259min & 82.1\% & 88.8\% \\
 & MF & 1 & 663 & 641,032 & 214min  & 81.8\% & 88.8\% \\ [1ex]
\\
\pptn{} (49) & 3SF & 1 & 577 & 327,908 & 32min  & 77.6\% & 89.8\% \\
 & MF & 1 & 393 & 327,908 & 27min  & 83.7\% & 89.8\% \\ [1ex]
\bottomrule  \\
\end{tabular}
\caption{Results in the ADDGT setting. The columns denote the method of substructure generation (SG), the number of representatives (\#REP) used for selecting substructures, the average number of candidates (\#CAND) and time taken for each instance, and the percentage of instances with ground truth ranked in the top 5 or 20 instances. The average time is based on running each test case on 4 parallel threads.}
\label{tab:easy}
\end{table}
\begin{table}
\centering
\small
\begin{tabular}{lcrrrrrr}
\toprule
Class (\#cases) & \#REP & \#SSTR & \#CAND 	& Top 5 	& Top 20 & \#G20 & Micro-F1 \\
\midrule
\ppn{}(285) 	&  1 & 397 	& 291,319 	& 43.5\% 	& 54.4\% 	&  36.5\% & 91.5\%	  \\
				&  2 & 885 	& 291,319 	& 53.7\% 	& 63.9\% 	& 26.7\% & 92.6\%  \\
				&  3 & 	1550		& 291,319 	& 62.8\%	& 70.9\%	&  20.4\% & 91.2\% \\
\pptn{} (49)  	&  1 & 160 	& 84,424	& 73.5\% 	& 81.6\% & 4.1\% & 91.9\%  \\
				&  2 & 343 	& 84,424 	& 77.6\% 	& 87.8\%& 2.0\% & 92.7\%  \\
\bottomrule \\
\end{tabular}
\caption{Results in the STRICT setting. The columns denote the number of representatives (\#REP) used for selecting substructures, the average number of candidates (\#CAND), the percentage of instances with ground truth ranked in the top 5 or 20 instances, the percentage of instances where the group (of candidates with equal scores) containing the ground truth has a size bigger than 20, and the micro-F1 of the substructure classifiers.}
\label{tab:hard}
\end{table}

For the STRICT settings, we only experiment with the MF substructure generation approach. As shown in Table~\ref{tab:hard}, our approach could still rank more than 80\% of cases in the top 20 for the smaller compound class of \pptn{}, but only 70.9\% for \ppn{} when 3 representatives were used to generate candidates for substructure selection. In Table~\ref{tab:hard}, we see that the micro-F1 for both \ppn{} and \pptn{} are similar. The reason for the worse performance in \ppn{} is due to the fact that substructure selection with 1 to 3 representatives does not provide enough substructures, and resulted in large equivalence classes of candidates with equal scores. When this happens, we take the worst ranking of the group containing the ground truth as the rank of the ground truth. Performance improves as we increase the  number of representatives used for substructure selection, resulting in more substructures used for discriminating between candidates. In the column \#G20, we show the percentage of instances with more than 20 structures in the equivalence class (of structures with equal scores) of the ground truth: these are cases where the approach fails automatically to rank the candidate in the top 20. We see that the majority of mistakes made (e.g., 45.6\%, 36.1\% and 29.1\% for \ppn{} for 1, 2 and 3 representatives) are due to the fact that the group of the ground truth contains more than 20 candidates (E.g., 36.5\%, 26.7\% and 20.4\% respectively). 

We also experimented with support vector machines~\cite{svm11} in scikit-learn~\cite{sklearn11} with worse  results. For example, for SVM with linear kernel, we could only achieve 69.4\% for \pptn{} with 2 representatives, compared to 87.8\% with neural networks. 

We show the  average run times for ranking the candidates generated by MOLGEN in Table~\ref{tab:easy}, where the run time is computed by running 4 parallel threads on each candidate, on a  Intel(R) Xeon(R) CPU E5-2620 v4 at 2.10GHz. This timing, however, does not include the time MOLGEN takes to generate candidates. In most cases, MOLGEN can finish generating all candidates in less than 1 hour.

\subsection{Analysis and Discussion}

\begin{figure}
\centering
\begin{subfigure}{0.24\textwidth}
\includegraphics[width=2.2cm]{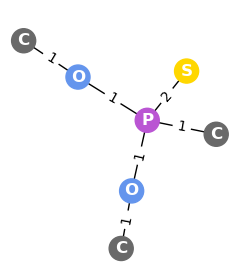}
\caption{0.9997}
\label{fig:anala}
\end{subfigure}
\begin{subfigure}{0.24\textwidth}
\includegraphics[width=2.2cm]{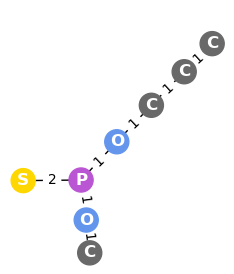}
\caption{0.9819}
\label{fig:analb}
\end{subfigure}
\begin{subfigure}{0.24\textwidth}
\includegraphics[width=2.2cm]{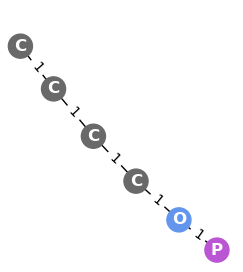}
\caption{0.1589}
\label{fig:analc}
\end{subfigure}
\begin{subfigure}{0.24\textwidth}
\includegraphics[width=2.2cm]{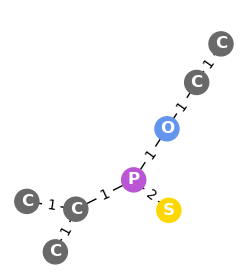}
\caption{0.0011}
\label{fig:anald}
\end{subfigure}
\caption{Example fragments with their predicted scores for O,O-dipropyl methylphosphonothionate, with the probability of the prediction that they are present in the structure given the mass spectrum.}
\label{fig:analysis1}
\end{figure}

In order to better understand this prediction model, we take a look at the predicted substructures for some of the test cases. As mentioned in Section~\ref{sect:rw}, the substructures used mat not be actual spectral fragments. Herein, we would take a look at 2 test cases, O,O-dipropyl methylphosphonothionate and O-isopropyl O-trimethylsilyl isopropylphosphonothionate, to understand the factors that affects the prediction accuracy.  
Looking at the \pptn{} results under the STRICT setting with 1 representative, the test case O,O-dipropyl methylphosphonothionate has the ground truth ranked as the most probable compound when its spectrum was analyzed. Substructures with –OCCC and –PC moieties were predicted as present (Figures~\ref{fig:anala} and~\ref{fig:analb}) while those with longer alkoxy and alkyl chains were predicted as absent (Figures~\ref{fig:analc} and~\ref{fig:anald}). This example showed that the machine learning is able to predict with high accuracy, the presence of fragments with the correct moieties.    

\begin{figure}
\centering
\begin{subfigure}{0.24\textwidth}
\includegraphics[width=2.2cm]{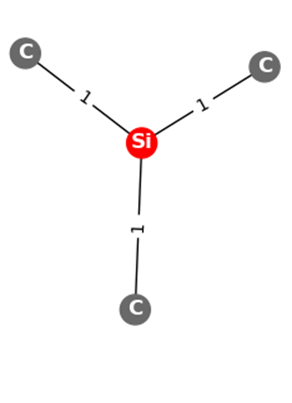}
\caption{0.9993}
\label{fig:anal7a}
\end{subfigure}
\begin{subfigure}{0.24\textwidth}
\includegraphics[width=2.2cm]{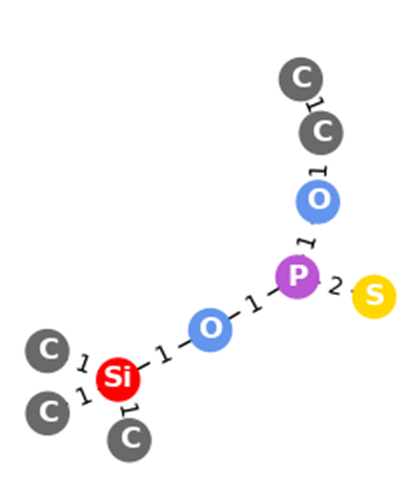}
\caption{0.9985}
\label{fig:anal7b}
\end{subfigure}
\caption{Example fragments with their predicted scores for O-isopropyl O-trimethylsilyl isopropylphosphonothionate, with the probability of the prediction that they are present in the structure given the mass spectrum in the two representative setting.}
\label{fig:analysis2}
\end{figure}

As we compare the PPTN results under the STRICT setting that made use of both 1 and 2 representatives, the test case O-isopropyl O-trimethylsilyl isopropylphosphonothionate has the ground truth ranked $3842^{nd}$ and $8^{th}$ positions respectively. While the STRICT setting with 1 representative used a dialkyl alkylphosphonothionate, the STRICT setting with 2 representatives used both a dialkyl alkylphosphonothionate and silylated alkyl alkylphosphonothionate. The addition of a silylated representative resulted in Si-containing differentiating fragments in the second setting. As we take a look at the fragments, the trimethylsilyl moiety (Figure~\ref{fig:analysis2}) is indeed predicted with high confidence to be present in the test case and this resulted in a huge improvement in the ground truth ranking. Therefore, we should choose representatives to contain the different functional groups that the model would be trained to recognise. 

\section{Conclusion}

In this paper, we proposed to automate the chemical  elucidation process to produce a ranked list of structures given  the mass spectrum and chemical formula of a compound.  We conducted experiments within compound classes in the OPCW VGWD database. Since our experiments showed we could classify instances into compound classes accurately (see Section~\ref{sec:ccclass}), this does not compromise the applicability of our approach. We experimented with two compound classes of chemical agents from the OPCW VGWD database,  and showed that we can rank the correct structure in the top 20 positions in over 88\% of the test cases for the \pptn{} compound class, and over 71\% for the \ppn{} class.  The computational time taken to rank the candidates of one test case is usually under one hour. If the correct structure is ranked within the top 5 or 20 candidates, the chemist can verify these structures experimentally or by computational mass spectral simulation to confirm the correct structure. In future work, we would like to investigate structure generation approaches such as~\cite{jin18,li18,simonvsky18} to bypass MOLGEN in this workflow. We believe that this automated workflow would be important in improving the efficiency of the work of analytical chemists.

\section*{Acknowledgements}
\vspace{-2mm}

We would like to thank Kian Ming Adam Chai, Hoe Chee Chua and Lee Hwi Ang for useful discussions.

\setlength{\bibsep}{2pt} 


\end{document}